\documentclass[12pt,preprint]{aastex}
\usepackage{graphicx}
\usepackage{graphics}

\slugcomment{}

\shorttitle{}
\shortauthors{} 

\begin{document}

\title{Close Binaries with Infrared Excess: Destroyers of Worlds?}

\author{M.~Matranga, J.J.~Drake, V.L.~Kashyap}
\affil{Harvard-Smithsonian Center for Astrophysics, Cambridge, MA 02138}

\author{M.~Marengo}
\affil{Dept. of Physics and Astronomy, Iowa State University, Ames, IA 50011}
   
\author{M.J.~Kuchner}
\affil{NASA Goddard Space Flight Center, Exoplanets and Stellar Astrophysics Laboratory Greenbelt, MD 20771}


\begin{abstract}

We present the results of a {\it Spitzer} photometric investigation
into the IR excesses of close binary systems.  In a sample of 10
objects, 
excesses in IRAC and MIPS24 bands  implying the presence of warm dust are found for 3.  For 2 objects we
do not find excesses reported in earlier IRAS studies.  We discuss
the results in the context of the scenario suggested by Rhee and
co-workers, in which warm dust is continuously created by destructive
collisions between planetary bodies.   
%
A simple
numerical model for the steady-state distribution of dust in one IR excess
system shows a central clearing of radius 0.22~AU caused by dynamical perturbations
from the binary star.  This is consistent with the size of the central clearing
derived from the Spitzer spectral energy distribution. We conclude that
close binaries could be efficient ``destroyers of worlds'',  lead
to destabilize the orbits of their planetary progeny by magnetically-driven
angular momentum loss and secular shrinkage of the binary separation.

\end{abstract}


\section{Introduction}

Perhaps the most spectacular discovery of the {\it Infrared
Astronomical Satellite (IRAS)} was the detection around a small
handful of normal main-sequence stars of excess infrared (IR)
radiation from remnant disks now understood to represent debris
related to planet formation (e.g.\ \citealt{Aumann.etal:84}).  
These disks are largely dissipated in 
in a few hundred Myr by collisional degradation of larger
bodies combined with Poynting-Robertson (PR) and stellar wind drag,
radiation pressure and evaporation \citep{Aumann.etal:84}. 
Recent {\it Spitzer} surveys at 70~$\mu$m find a disk fraction of 
$\sim 10$\%\ for F, G and K stars with ages up to 3~Gyr or so  \citep{Bryden.etal:06,Hillenbrand.etal:08}.


One other group of
stars with IR excesses has remained inconspicuous. The RS~CVn class of
tidally-interacting close binaries, in which at least one component has generally reached the sub-giant evolutionary phase \citep{Hall:76}, 
were discovered by IRAS to have an IR excess
occurrence rate of about 40\% ---twice as high
as main-sequence stars and 3--4 times higher than giants
(\citealt{Busso.etal:87,Busso.etal:88,Scaltriti.etal:93}, and references
therein).  The observed excesses typically amounted to several percent
of the stellar bolometric luminosity and corresponded to a blackbody
temperature of order 1500~K.  Within current scenarios explaining IR
excesses around late type stars, RS~CVns should not have any: they
represent a significantly older population than the few 100~Myr
dusty disk lifetime \citep[e.g.][]{Eker:92}, and have not undergone
any evolutionary phase during which substantial mass loss and dust
formation is expected.  Despite this puzzle, the IR excess RS~CVns
have received scant attention since their initial discovery.

More recently, \citet{Weinberger:08} found BD~+20~307, ``the
dustiest known main-sequence star'' \citep{Zuckerman.etal:08}, to be
a close spectroscopic binary with a period of 3.4 days and estimated
its age to be greater than 1~Gyr.  The parallel between the 3.4 day
binary BD~+20~307 and the IR excess RS~CVns---spectroscopic binaries
with periods of a few days---is striking.  The dust emission of
BD~+20~307 amounts to an unusually large 4\% of its bolometric
luminosity, and has an unusually high temperature of $\sim 650$~K
\citep{Song.etal:05}.  \citet{Song.etal:05}, \citet{Rhee.etal:08}, and 
\citet{Zuckerman.etal:08} suggest the warm dust in BD~+20~307 likely arises
from recent pulverising collisions between planets or planetesimals.  
That such close binaries can harbor planetary systems has been demonstrated by 
the planet around  HW~Vir (\citealt{Lee.etal:09}).

Here we present a {\it Spitzer} photometric survey of a
small sample of 10 RS~CVns to probe deeper into the purported IR
excesses of this class.





\section{Observations and Data Analysis}
\label{s:obsanal}

The data presented in this paper were obtained by the {\it Spitzer
Space Telescope}  \citep{Werner.etal:04,Gehrz.etal:07} as part of the 
General Observing program PID 20713.
{\it Infrared Array Camera} (IRAC;
\citealt{Fazio.etal:04b}) observations were made  in High Dynamic Range mode 
between 2005 Jul and Dec.  Targets were observed in 3.6, 4.5, 5.8 and 8.0~$\mu$m bands 
for 12s in each position of a 12 point Reuleaux pattern dither map, for a total effective integration time of 124.8s. 
{\it Multiband Imaging Photometer} (MIPS; \citealt{Rieke.etal:04}) observations were obtained in the 24, 70 and 160~$\mu$m bands in photometry mode between 2005 Nov and 2007 Jan, 
with an exposure time of 10s, repeated for 2 cycles
(24$\;\mu$m) or 3 cycles (70 and 160~$\;\mu$m).  Reduced image mosaics were produced with the IRACproc  custom software \citep{Schuster.etal:06}, while for MIPS 
we used the Spitzer Science Center pipeline mosaics.  
These were analysed using standard aperture (MIPS) and Point spread Function (PSF) based (IRAC) photometry, the latter adopting the procedure described in \citet{Marengo.etal:09}, that enable deriving the flux of core-saturated sources. 
All photometry results are reported in Table~\ref{t:photo}.
 
Only one source (UX Ari) was detected at $70\;\mu m$ at $\geq 3 \sigma$.  For two other sources (CF Tuc and II Peg) we found $2 \sigma$ level signals.  For all other sources we report $3 \sigma$ upper limits.
The well known MIPS 160 micron near-IR spectral leak prevented us from using data taken in this band.

Our analysis procedure was similar to that followed by \citet{Busso.etal:87,
Busso.etal:88,Scaltriti.etal:93} and involved comparison of stellar model
fluxes, fitted to observed optical and near-IR photometry, with the 
{\it Spitzer} IR photometry.
For each binary listed in Table~\ref{t:photo}, we modelled
\emph{BVRIJHK} photometry culled from literature with a spectral energy
distribution (SED) model comprising the sum of two Kurucz photospheric
models\footnote{http://wwwuser.oat.ts.astro.it/castelli/grids.html} (one for
each stellar component in the case of double-lined spectroscopic binaries, 
weighted by stellar radius and effective temperature according to 
$r^2T_{\rm eff}^4$). Since surface
spots resulting from magnetic activity on RS~CVns can give rise to
detectable secular photometric changes (thus leading to some having
been designated as variables), we adopted photometric uncertainties
that generally encompassed observed ranges of variations for each
star.  For dwarfs stars, we adopted the
effective temperature--radius relation of \citet{Gray:92} with initial values for
effective temperature from the literature; for more evolved
components we culled values for temperature, radius and/or mass and 
surface  gravity from the literature. Final radius and effective temperature model parameters for each system were honed by exploring 
the grid of Kurucz models using a $\chi^2$ minimisation. 

\begin{figure}
\epsscale{0.55}
\plotone{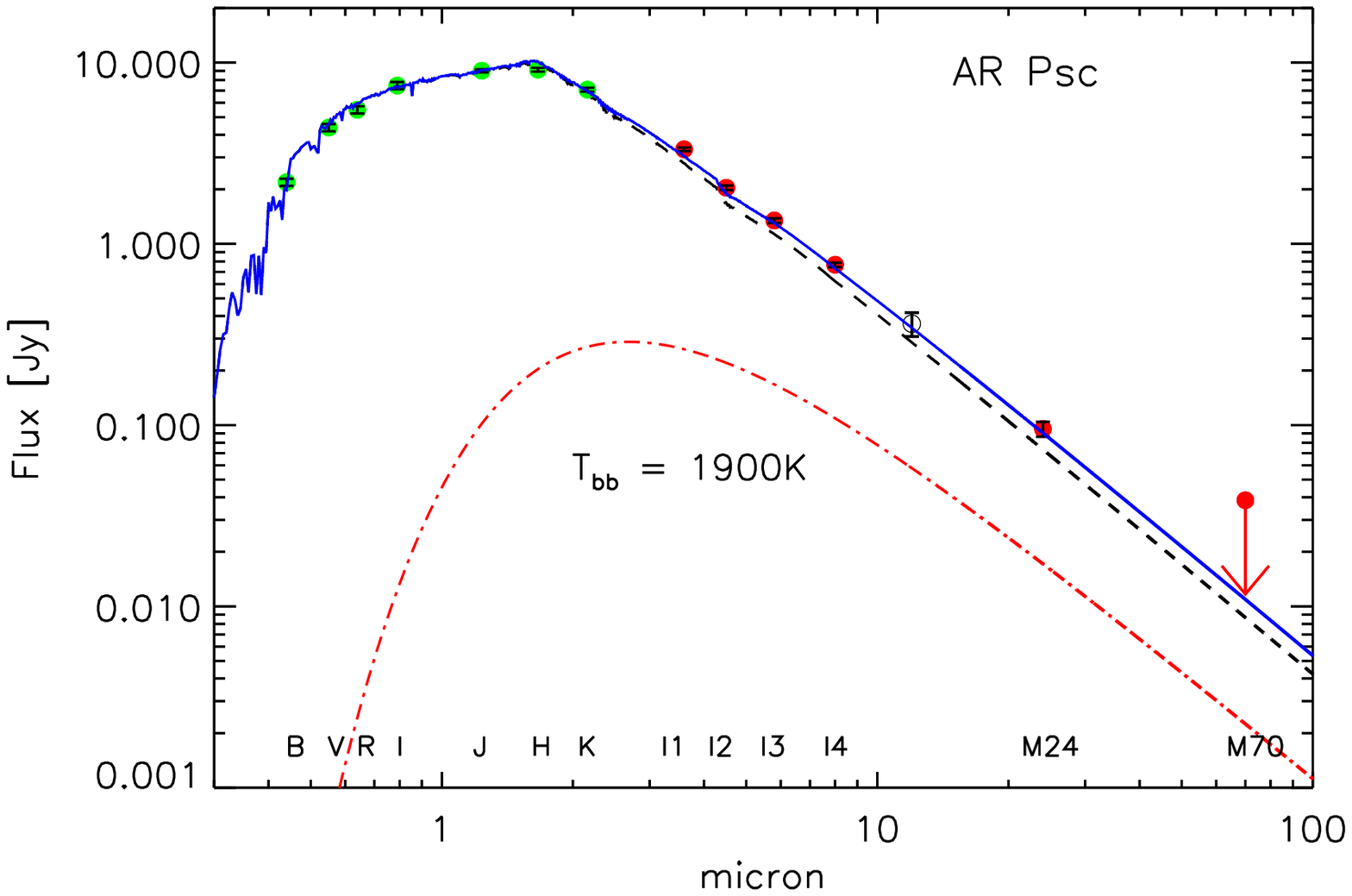}
\plotone{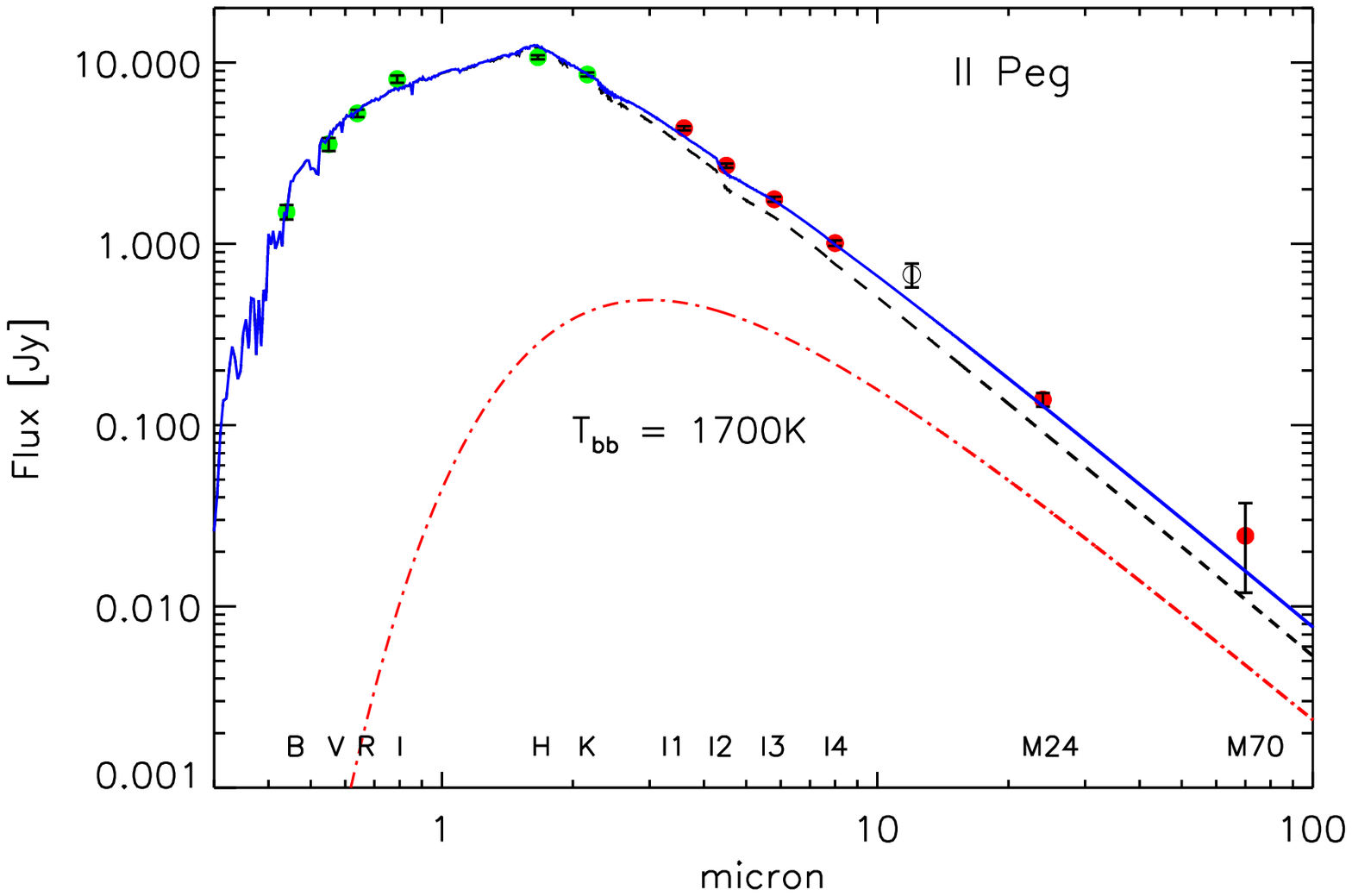}
\plotone{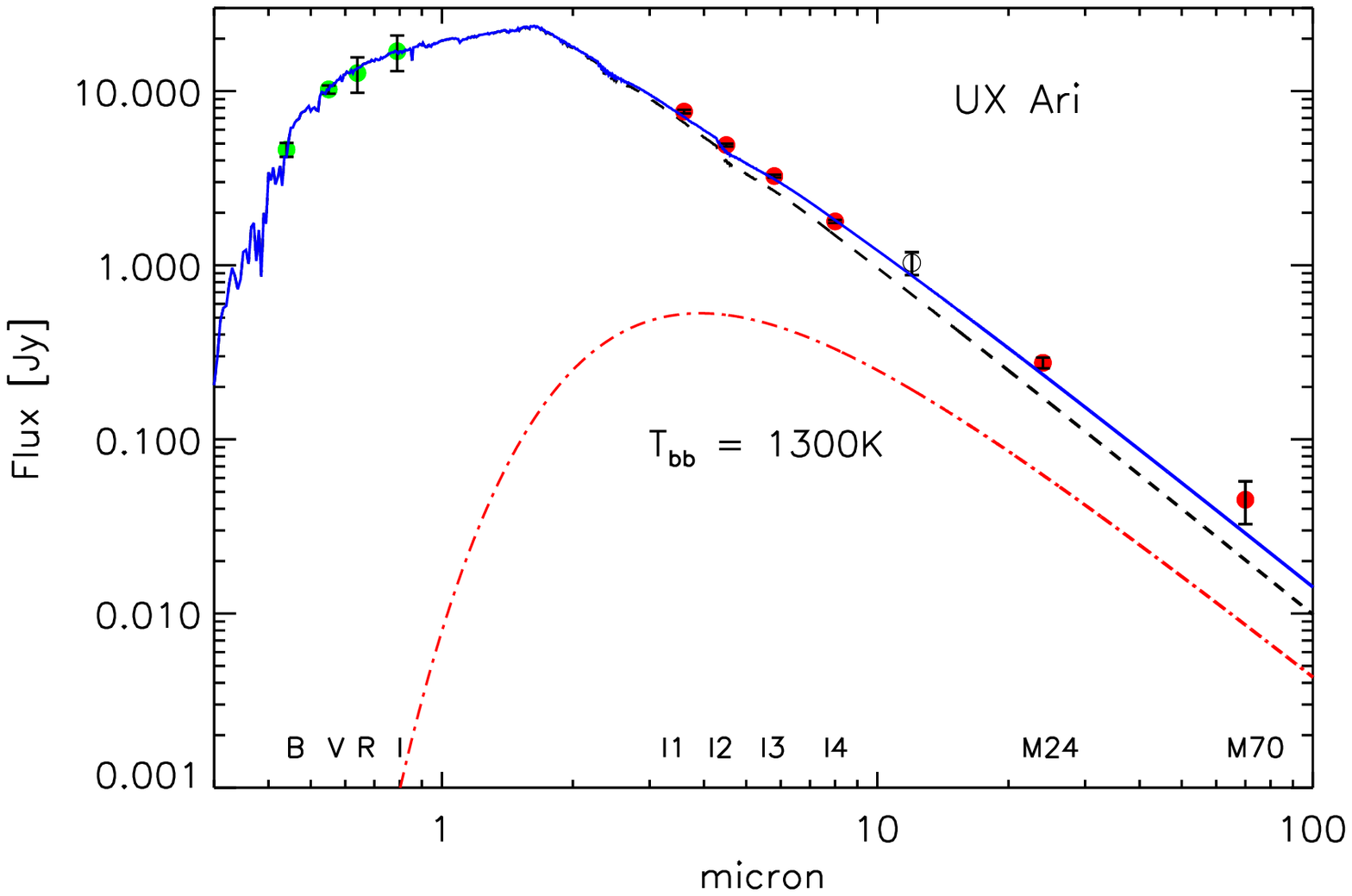}
\caption{\footnotesize Observed and model SEDs for the IR excess stars 
found in this survey.  The black (dashed) line represents the photospheric
component, the red (dot-dashed) line corresponds to the blackbody dust component
and the blue (solid) line represents the sum of photospheric and blackbody dust models.
Photometric measurements are represented by different coloured
symbols: green - $BVRIJHK$, red - IRAC (I1, I2, I3, I4) and MIPS (M24, M70). IRAS $12\,\mu m$ data are indicated by hollow black circles.  \label{f:seds}}
\normalsize
\end{figure}

Final stellar parameters are listed in 
Table~\ref{t:fit}.
Comparison of adopted Kurucz models with {\it Spitzer} photometry
revealed 
satisfactory fits for all but the three sources: AR~Psc, II~Peg and UX~Ari.  For
these systems, no combination of Kurucz photospheres 
could reproduce the measured IRAC and MIPS fluxes: in all three 
cases, the observed IR fluxes were
too high compared to the predicted model fluxes.  
Including a blackbody as an additional component, fitting the
temperature and normalization as free parameters (see Table~\ref{t:fit}), greatly improved the SED model fits.   Results are illustrated in 
Figure~\ref{f:seds}), together with IRAS Faint Source Catalog v2.0 
$12\, \mu m$ fluxes\footnote{http://irsa.ipac.caltech.edu}.
In agreement with earlier studies (e.g., \citet{Scaltriti.etal:93}), the best-fit
blackbody temperatures lie in the range 1300-1900~K with an uncertainty of
200--300~K.  The low-level 70~$\mu m$ emission in II Peg is consistent with the long wavelength tail of the hot blackbody excess, while that for UX~ARI hints at the possible  presence of additional cold dust.  
For other binaries that do not show a detectable IR excess, we have
estimated $3\sigma$ upper limits.  Assuming a
nominal blackbody temperature of 1500~K, we find that for these
systems the blackbody luminosity must be less than $\approx2-2.5$\%
of the bolometric luminosity (see Table~\ref{t:photo}).

To estimate more realistic dust temperatures, non-blackbody circumstellar dust with an emissivity law of the type 
$\epsilon=1, \, \lambda \leq \lambda_0; \, \epsilon=(\lambda_0/\lambda)^q, \, \lambda > \lambda_0$, where $\lambda_0$ corresponds 
approximately to the grain size and $q$ is a parameter of order unity, depending 
on grain type, as described by, e.g., \citet{Backman.Paresce:93} were used.  
In general, fits 
using a grain emissivity law (Table~\ref{t:fit}) required lower temperatures.  For both II~Peg and AR~Psc, the best-fit values of $\lambda_0$ and $q$ are $3\;\mu$m and 1  respectively, whereas for UX~Ari these parameters were not well-constrained by the data.

We note two systems in Table~\ref{t:photo} for which we find no significant excess 
at levels below earlier IRAS-based detections; either the earlier detections were spurious, or the excesses have waned in the intervening years.

\section{Discussion}
\label{s:disc}

The span from 3.6--24~$\mu$m covered by our {\it Spitzer} IRAC and
MIPS photometry is very effective for diagnosing the presence of warm
and even hot dust.  The requirement for a purely stellar SED is that
the IR photometric indices lie along the Rayleigh-Jeans tail of the
photospheric IR (essentially blackbody) continuum.  The SEDs
illustrated in Figure~\ref{f:seds} fail to do this in the sense that
the observed slope is incorrect: the departure from stellar SEDs
increases toward longer wavelengths.  From the blackbody fits, 
we have determined the fraction of the stellar bolometric
luminosities reprocessed into IR emission and list these in
Table~\ref{t:photo}: we find values ranging from 0.7 to 1.9\%.

Our detection of IR excess in the spectrum of II~Peg confirms the
earlier photometric detection by \citet[][see also
\citealt{Scaltriti.etal:93}]{Busso.etal:87} and the preliminary {\it
ISOPHOT} spectrum obtained by \citet{Rodono.etal:98}.  Of our other
two detections, UX~Ari was found to have a substantial near-IR
excess in $JHK$ photometry by \citet{Verma.etal:83} but was not found
to have a significant excess in IRAS photometry by
\citet{Busso.etal:88}. AR~Psc instead represents a new IR excess detection. 

As noted by \citet{Scaltriti.etal:93}, the observed IR excesses cannot be fitted with a standard free-free emission spectrum from, e.g., an ionized wind, whose $\nu^{0.6}$ frequency dependence \citep[e.g.][]{Wright.Barlow:75} is much too flat.  Moreover, the observed fluxes would imply mass loss rates of order $10^{-8}$~$M_{\odot}$~yr$^{-1}$---orders of magnitude higher than expected for these stars.   
\citet{Busso.etal:87,Busso.etal:88} and  \citet{Scaltriti.etal:93} interpreted the
observed excesses in terms of dust shells.  While the implied dust
temperatures are high in the context of dust in debris disks, we
emphasise that unresolved low-mass stars or brown dwarfs cannot
explain the excesses because typical luminosities of these objects 
for the required temperatures are $L < 10^{-3}\; L_{\odot}$
\citep{Riddick.etal:07} which is more than an order of
magnitude lower than observed.

\citet{Busso.etal:87} and \citet{Scaltriti.etal:93} 
suggested the dust arises from the
accumulation of stellar winds driven by magnetic activity.  However,
in order to reach temperatures of $\sim 1200-1400$~K, dust
must reside close to the central stars at typical distances of $\sim
0.1-0.3$~AU .  Dust in such a location
is short-lived and any grains that form will be rapidly
dissipated by PR drag.
\citet{Rhee.etal:08} estimate the lifetimes of larger grains of a few
$\mu$m in size around BD~+20~307 to be only a few decades.
Minimum dust masses able to explain the excesses of II~Peg, AR~Psc and UX~Ari are $1.3\times 10^{-12}$, $2.8\times 10^{-13}$ and $4.7\times 10^{-13}$~$M_{\odot}$, respectively; a standard gas:dust ratio implies total masses 100 times higher.
\citet{Wood.etal:05} found very active stars might lose mass at rates perhaps 
up to $10^{-12}M_\odot$~yr$^{-1}$ (100 times that of the Sun), meaning 10-100 years of accumulated wind would be necessary to explain the dust.  It
seems highly unlikely that such fully-ionized, magnetically-driven winds and coronal mass
ejections with outflow velocities of several hundred km~s$^{-1}$ would accumulate in this way.

We are lead to the suggestion of \citet[][see also 
\citealt{Weinberger:08,Zuckerman.etal:08}]{Rhee.etal:08} that such 
warm dust originates from collisions between planetary bodies.  
Since grains produced in this way will be dissipated relatively quickly, such  
collision events must be recent.  

The problem, then, is why stars of such ages 
as the RS~CVn-type binaries should be producing dust when debris disks around
single stars and wide binaries are largely dissipated in a few hundred Myr.
In the case of BD+20~307, \citet{Zuckerman.etal:08} noted possible systemic
radial velocity variations and suggested an unseen brown dwarf or low-mass star
could have destabilized the orbits of rocky bodies.  Such destabilization,
occurring many millions of years after a ``first generation'' debris disk would have been mostly dissipated,
could stimulate new dust production in Gyr old systems, reviving whatever debris 
disk remains around these stars.

We suggest another possible mechanism resulting from 
the conspicuous property of tidally-interacting binaries not shared by their 
non-interacting single and binary brethren of similar age: copious magnetic
activity manifest in the form of high levels of surface spots, chromospheric 
and X-ray emission.  Such activity is thought responsible for the rotational
braking through angular momentum loss of main-sequence stars from zero-age onwards
\citep[e.g.][]{Mestel:84}.
Angular momentum loss of the dwarf precursors to RS~CVns, such as BD+20~307, as well as the more X-ray bright RS~CVns themselves, results in substantial orbital shrinkage (e.g. by factors of up to 2 or more) through 
spin-orbit tidal coupling on timescales of hundreds of Myr to a Gyr \citep[e.g.][]{Eggleton.Kiseleva-Eggleton:02}.  This change in the orbit
causes proportional changes in the locations of both mean motion
resonances and secular resonances in whatever planetary system the binary has. 
We postulate that the sweeping of these resonances can lead to both destabilization 
of the orbits of small bodies and resonant capture; both of these processes can 
increase the collision rate among these bodies.  This mechanism would provide a more general means of stimulating new eras of copious dust production and debris disk revival in Gyr old systems without the need to invoke perturbations from additional companions.

One potential problem for a collisional origin of the dust is the high frequency of detection---40\%\ according to \citet{Busso.etal:88} and \citet{Scaltriti.etal:93}.  Our pilot study sample is partially biased toward systems with previously identified excesses.  If our non-detection of excesses in CF~Tuc and WY~Cnc imply earlier detections were spurious, the excess frequency implied by this and our detection of one excess in five objects not previously studied is $\sim 20$\%.   An evolutionary timescale of 1~Gyr then implies an integrated warm dust production period---whether in a single or multiple episodes---of 200~Myr.   For a 100~yr dissipation timescale of $10^{-12}$~$M_{\odot}$ of dust, the total mass required is $\sim 2\times 10^{-6}$~$M_{\odot}$ or about 1 Earth mass, consistent with a planetary origin.
 
\subsection{Dynamical Debris Disk Model}
\label{s:model}

To examine the interaction between a close binary star and a debris disk
we ran a numerical model \citep[see, e.g.][]{Wilner.etal:03} of a dust cloud around II~Peg using the dust properties
consistent from our {\em Spitzer} photometry.  II~Peg was chosen because it is the best studied of the three systems for which we detect IR excesses:  we assumed masses of 1.67 and $0.23 M_{\odot}$, a semi-major 
axis of 0.087~AU, eccentricity of 0.0, and total luminosity of $4.87L_\odot$ based on the stellar parameters reported by \citet{Ottmann.etal:98} and \citet{Frasca.etal:08}.

\begin{figure}
  \begin{center}
   \includegraphics[width=6in]{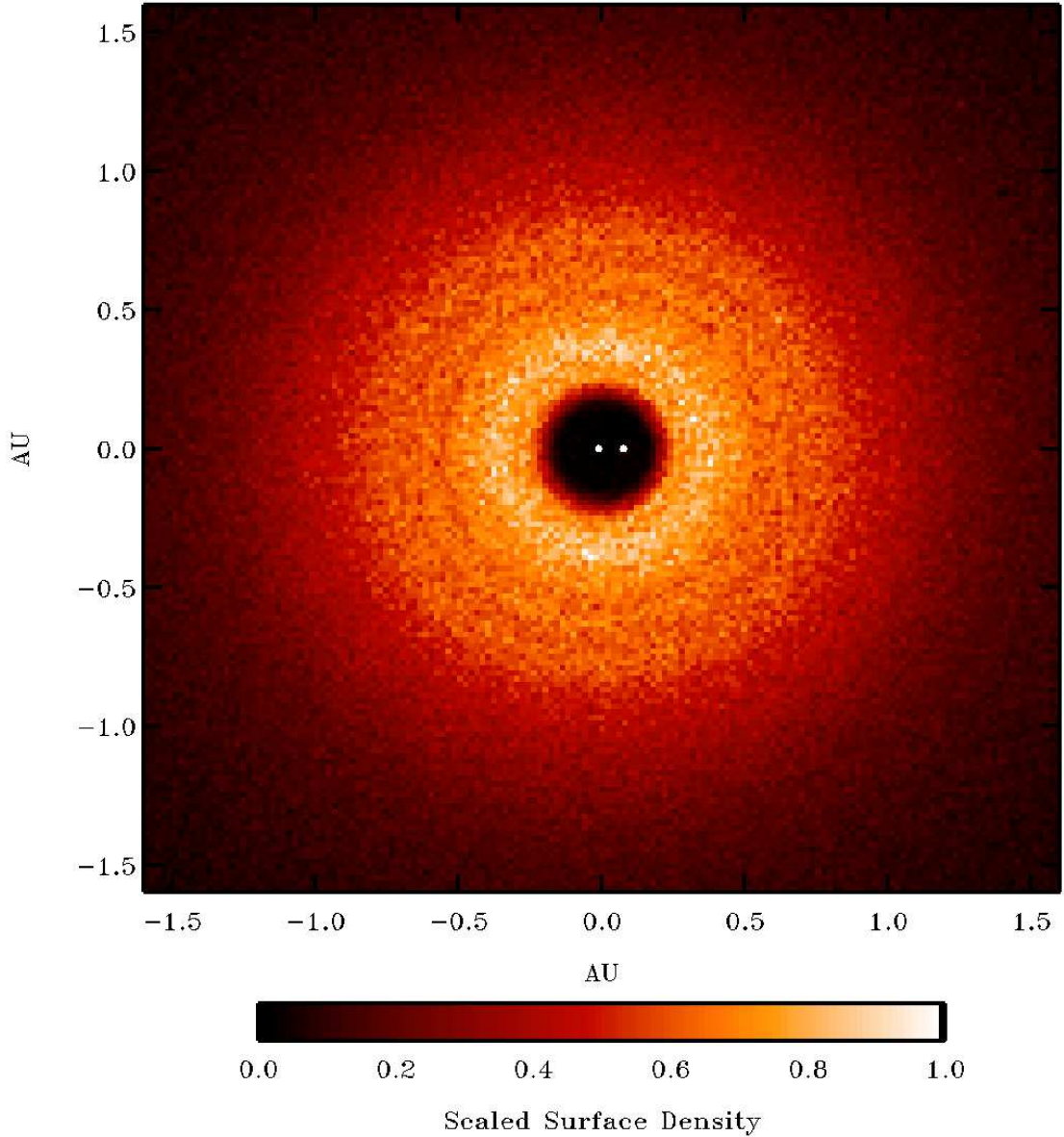}
  \end{center}
  \caption{The steady-state surface density distribution of a dust cloud model
  for II~Peg. See Section 3 for details. White dots indicate the locations of the two stars. The dust surface density distribution is azimuthally symmetric and shows a central clearing centered on the barycenter of the binary. The radius of this clearing in the dynamical model is 0.22~AU, which is consistent with grain temperatures derived from the observed spectral energy distribution.}
  \label{f:nummod}
\end{figure}

For this initial model, we focus on dynamical clearing and neglect grain-grain collisions, which can also alter the disk morphology \citep[e.g.][]{Stark.Kuchner:08}.
Dust was released from 1~AU with initial eccentricities and inclinations uniformly distributed over the ranges 
0--0.1 and 0--0.05, respectively, with the rest of the orbital elements uniformly distributed over 0--$2 \pi$.
We numerically integrated the orbits
of 620 dust grains, subject to gravity from the stars, radiation pressure, and PR drag, until the particles either collided with one of the stars or was ejected beyond 4~AU. 
Another parameter in these models is $\beta$, the radiation pressure force on the grain divided by the gravitational force. 
In the vicinity of a star with the luminosity of II~Peg, a 3$\;\mu$m radius silicate sphere consistent with our models for the photometry corresponds to $\beta=0.463$.  We chose this value, and neglected the slight radiation pressure from the faint secondary star; we leave this detail for future models.
At regular time intervals, we accumulated the particle positions in a histogram, in the manner of many previous authors \citep[e.g.][]{Wilner.etal:03, Stark.Kuchner:08}.  This histogram models the steady-state surface 
density of the cloud, which is illustrated in Figure~\ref{f:nummod}.

Figure~\ref{f:nummod} shows a disk with a central clearing, centered 
on the barycenter of the binary.
White dots indicate the locations of the two stars. The dust surface density distribution is azimuthally symmetric, consistent with the analytic predictions of 
\citet{Kuchner.Holman:03}. There are 2--3 ring-like density enhancements, which might be artifacts of the narrow particle size distribution in these initial models, but no signs of clumps or spiral features. For interpreting the {\em Spitzer} data, the most important result is the size of the central clearing.  The radius of this clearing in the dynamical model is 0.22~AU and similar to the radius at which first order mean motion resonances overlap (and actually coincident with the 4:1 mean motion resonance).  This is consistent with our models for the spectral energy distribution for which the grain emissivity law temperature implies a radius $0.24_{-0.06}^{+0.1}$~AU.  A dust sublimation temperature of 1700~K implies that the dust can survive at 0.22~AU, and that the clearing radius is determined by the system dynamics, rather than grain evaporation. 

In Section~\ref{s:obsanal}, we noted that IRAS excesses for two systems in our {\it Spitzer} sample are not confirmed. 
The Poynting-Robertson time for 3~$\mu$m grains 
is \citep{Wyatt.Whipple:50} $T_{PR}=400{(a_0/1AU)^2}L_{\odot}/\beta L_{\star}$~yr. For the grains in the II~Peg model, $a_0=1 AU, L_{\star}=4.87 L_{\odot}, \beta=0.463$, and the PR time is 177 years.  But if the grains are launched near the star at 0.3 AU instead of 1~AU, the PR time drops to 16 years---sufficiently short for IRAS epoch dust to have dissipated by the time {\it Spitzer} observations were made.  Either the earlier detections were spurious, or excesses could have waned, suggestive of dust being produced in a succession of discrete events.

\section{Conclusions}

A {\it Spitzer} survey of ten tidally-interacting binary systems finds IR excesses for three. 
The excesses are consistent with significant amounts of warm 
dust close to the central stars.  
A numerical model for the steady-state distribution of dust in II~Peg finds 
a clearing radius of 0.22~AU, consistent with the observed optical-IR SED,
and suggests the warm dust temperatures result from inward migration 
of dust due to PR drag and central clearing caused by 
dynamical perturbations.

Reported IRAS excesses for CF~Tuc and WY~Cnc are not confirmed by {\it Spitzer}, indicating a possible waning of dust in the last few decades.  
This is consistent with PR drag, for which we find the timescale for dust depletion around II~Peg could be as short as a decade or so, suggesting ongoing dust production.  We 
echo the suggestion of  \citet{Rhee.etal:08} that dust could arise from collisions of planetary bodies, and speculate that these old binaries can enter epochs in which 
they destabilize the orbits of their planets through shrinkage of their own binary 
orbit due to magnetically-driven angular momentum loss.


\acknowledgments

M. Matranga was supported by {\it Spitzer} Contract 1279130; JJD and VLK were funded by NASA contract NAS8-39073 to the {\it Chandra} X-ray Center.   Observations were 
made  with {\it Spitzer}, operated by JPL under NASA contract 1407.


\bibliographystyle{apj}

\begin{thebibliography}{47}
\expandafter\ifx\csname natexlab\endcsname\relax\def\natexlab#1{#1}\fi

\bibitem[{{Alekseev} \& {Kozhevnikova}(2004)}]{Alekseev.Kozhevnikova:04}
{Alekseev}, I.~Y., \& {Kozhevnikova}, A.~V. 2004, Astrophysics, 47, 443

\bibitem[{{Aumann} {et~al.}(1984){Aumann}, {Beichman}, {Gillett}, {de Jong},
  {Houck}, {Low}, {Neugebauer}, {Walker}, \& {Wesselius}}]{Aumann.etal:84}
{Aumann}, H.~H., {et~al.} 1984, \apjl, 278, L23

\bibitem[{{Backman} \& {Paresce}(1993)}]{Backman.Paresce:93}
{Backman}, D.~E., \& {Paresce}, F. 1993, in Protostars and Planets III, ed.
  E.~H. {Levy} \& J.~I. {Lunine}, 1253--1304

\bibitem[{{Bryden} {et~al.}(2006){Bryden}, {Beichman}, {Trilling}, {Rieke},
  {Holmes}, {Lawler}, {Stapelfeldt}, {Werner}, {Gautier}, {Blaylock}, {Gordon},
  {Stansberry}, \& {Su}}]{Bryden.etal:06}
{Bryden}, G., {et~al.} 2006, \apj, 636, 1098

\bibitem[{{Busso} {et~al.}(1988){Busso}, {Scaltriti}, {Origlia}, {Persi}, \&
  {Ferrari-Toniolo}}]{Busso.etal:88}
{Busso}, M., {Scaltriti}, F., {Origlia}, L., {Persi}, P., \& {Ferrari-Toniolo},
  M. 1988, \mnras, 234, 445

\bibitem[{{Busso} {et~al.}(1987){Busso}, {Scaltriti}, {Persi}, {Robberto}, \&
  {Silvestro}}]{Busso.etal:87}
{Busso}, M., {Scaltriti}, F., {Persi}, P., {Robberto}, M., \& {Silvestro}, G.
  1987, \aap, 183, 83

\bibitem[{{Byrne} {et~al.}(1998){Byrne}, {Abdul Aziz}, {Amado}, {Arevalo},
  {Avgoloupis}, {Doyle}, {Eibe}, {Elliott}, {Jeffries}, {Lanzafame}, {Lazaro},
  {Murphy}, {Neff}, {Panov}, {Sarro}, {Seiradakis}, \&
  {Spencer}}]{Byrne.etal:98}
{Byrne}, P.~B., {et~al.} 1998, \aaps, 127, 505

\bibitem[{{Cutispoto}(1991)}]{Cutispoto:91}
{Cutispoto}, G. 1991, \aaps, 89, 435

\bibitem[{{Cutispoto} {et~al.}(2001){Cutispoto}, {Messina}, \&
  {Rodon{\`o}}}]{Cutispoto.etal:01}
{Cutispoto}, G., {Messina}, S., \& {Rodon{\`o}}, M. 2001, \aap, 367, 910

\bibitem[{{Eggleton} \&
  {Kiseleva-Eggleton}(2002)}]{Eggleton.Kiseleva-Eggleton:02}
{Eggleton}, P.~P., \& {Kiseleva-Eggleton}, L. 2002, \apj, 575, 461

\bibitem[{{Eker}(1992)}]{Eker:92}
{Eker}, Z. 1992, \apjs, 79, 481

\bibitem[{{Fazio} {et~al.}(2004){Fazio}, {Hora}, {Allen}, {Ashby}, {Barmby},
  {Deutsch}, {Huang}, {Kleiner}, {Marengo}, {Megeath}, {Melnick}, {Pahre},
  {Patten}, {Polizotti}, {Smith}, {Taylor}, {Wang}, {Willner}, {Hoffmann},
  {Pipher}, {Forrest}, {McMurty}, {McCreight}, {McKelvey}, {McMurray}, {Koch},
  {Moseley}, {Arendt}, {Mentzell}, {Marx}, {Losch}, {Mayman}, {Eichhorn},
  {Krebs}, {Jhabvala}, {Gezari}, {Fixsen}, {Flores}, {Shakoorzadeh}, {Jungo},
  {Hakun}, {Workman}, {Karpati}, {Kichak}, {Whitley}, {Mann}, {Tollestrup},
  {Eisenhardt}, {Stern}, {Gorjian}, {Bhattacharya}, {Carey}, {Nelson},
  {Glaccum}, {Lacy}, {Lowrance}, {Laine}, {Reach}, {Stauffer}, {Surace},
  {Wilson}, {Wright}, {Hoffman}, {Domingo}, \& {Cohen}}]{Fazio.etal:04b}
{Fazio}, G.~G., {et~al.} 2004, \apjs, 154, 10

\bibitem[{{Frasca} {et~al.}(2008){Frasca}, {Biazzo}, {Ta{\c s}}, {Evren}, \&
  {Lanzafame}}]{Frasca.etal:08}
{Frasca}, A., {Biazzo}, K., {Ta{\c s}}, G., {Evren}, S., \& {Lanzafame}, A.~C.
  2008, \aap, 479, 557

\bibitem[{{Gehrz} {et~al.}(2007){Gehrz}, {Roellig}, {Werner}, {Fazio}, {Houck},
  {Low}, {Rieke}, {Soifer}, {Levine}, \& {Romana}}]{Gehrz.etal:07}
{Gehrz}, R.~D., {et~al.} 2007, Review of Scientific Instruments, 78, 011302

\bibitem[{{Gray}(1992)}]{Gray:92}
{Gray}, D.~F. 1992, {The observation and analysis of stellar photospheres.},
  ed. D.~F. {Gray}

\bibitem[{{Hall}(1976)}]{Hall:76}
{Hall}, D.~S. 1976, in Astrophysics and Space Science Library, Vol.~60, IAU
  Colloq. 29: Multiple Periodic Variable Stars, ed. W.~S. {Fitch}, 287--+

\bibitem[{{Heckert} {et~al.}(1998){Heckert}, {Maloney}, {Stewart}, {Ordway},
  {Hickman}, \& {Zeilik}}]{Heckert.etal:98}
{Heckert}, P.~A., {Maloney}, G.~V., {Stewart}, M.~C., {Ordway}, J.~I.,
  {Hickman}, A., \& {Zeilik}, M. 1998, \aj, 115, 1145

\bibitem[{{Hillenbrand} {et~al.}(2008){Hillenbrand}, {Carpenter}, {Kim},
  {Meyer}, {Backman}, {Moro-Mart{\'{\i}}n}, {Hollenbach}, {Hines}, {Pascucci},
  \& {Bouwman}}]{Hillenbrand.etal:08}
{Hillenbrand}, L.~A., {et~al.} 2008, \apj, 677, 630

\bibitem[{{Ibanoglu}(1989)}]{Ibanoglu:89}
{Ibanoglu}, C. 1989, \apss, 161, 221

\bibitem[{{{\.I}bano{\v g}lu} {et~al.}(2005){{\.I}bano{\v g}lu}, {Evren},
  {Ta{\c s}}, \& {{\c C}ak{\i}rl{\i}}}]{Ibanoglu.etal:05}
{{\.I}bano{\v g}lu}, C., {Evren}, S., {Ta{\c s}}, G., \& {{\c C}ak{\i}rl{\i}},
  {\"O}. 2005, \mnras, 360, 1077

\bibitem[{{Kjurkchieva} {et~al.}(2000){Kjurkchieva}, {Marchev}, \&
  {Ogloza}}]{Kjurkchieva:00}
{Kjurkchieva}, D., {Marchev}, D., \& {Ogloza}, W. 2000, \aap, 354, 909

\bibitem[{{Kjurkchieva} {et~al.}(2005){Kjurkchieva}, {Marchev}, {Heckert}, \&
  {Ordway}}]{Kjurkchieva.etal:05}
{Kjurkchieva}, D.~P., {Marchev}, D.~V., {Heckert}, P.~A., \& {Ordway}, J.~I.
  2005, \aj, 129, 1084

\bibitem[{{Kuchner} \& {Holman}(2003)}]{Kuchner.Holman:03}
{Kuchner}, M.~J., \& {Holman}, M.~J. 2003, \apj, 588, 1110

\bibitem[{{Lee} {et~al.}(2009){Lee}, {Kim}, {Kim}, {Koch}, {Lee}, {Kim}, \&
  {Park}}]{Lee.etal:09}
{Lee}, J.~W., {Kim}, S., {Kim}, C., {Koch}, R.~H., {Lee}, C., {Kim}, H., \&
  {Park}, J. 2009, \aj, 137, 3181

\bibitem[{{Marengo} {et~al.}(2009){Marengo}, {Stapelfeldt}, {Werner}, {Hora},
  {Fazio}, {Schuster}, {Carson}, \& {Megeath}}]{Marengo.etal:09}
{Marengo}, M., {Stapelfeldt}, K., {Werner}, M.~W., {Hora}, J.~L., {Fazio},
  G.~G., {Schuster}, M.~T., {Carson}, J.~C., \& {Megeath}, S.~T. 2009, \apj,
  700, 1647

\bibitem[{{Messina}(2008)}]{Messina.etal:08}
{Messina}, S. 2008, \aap, 480, 495

\bibitem[{{Mestel}(1984)}]{Mestel:84}
{Mestel}, L. 1984, in Lecture Notes in Physics, Berlin Springer Verlag, Vol.
  193, Cool Stars, Stellar Systems, and the Sun, ed. S.~L. {Baliunas} \&
  L.~{Hartmann}, 49--+

\bibitem[{{Milano} {et~al.}(1986){Milano}, {Mancuso}, {Vittone}, {Dorsi}, \&
  {Marcozzi}}]{Milano.etal:86}
{Milano}, L., {Mancuso}, S., {Vittone}, A., {Dorsi}, A., \& {Marcozzi}, S.
  1986, \apss, 124, 83

\bibitem[{{Ottmann} {et~al.}(1998){Ottmann}, {Pfeiffer}, \&
  {Gehren}}]{Ottmann.etal:98}
{Ottmann}, R., {Pfeiffer}, M.~J., \& {Gehren}, T. 1998, \aap, 338, 661

\bibitem[{{Rhee} {et~al.}(2008){Rhee}, {Song}, \& {Zuckerman}}]{Rhee.etal:08}
{Rhee}, J.~H., {Song}, I., \& {Zuckerman}, B. 2008, \apj, 675, 777

\bibitem[{{Riddick} {et~al.}(2007){Riddick}, {Roche}, \&
  {Lucas}}]{Riddick.etal:07}
{Riddick}, F.~C., {Roche}, P.~F., \& {Lucas}, P.~W. 2007, \mnras, 381, 1077

\bibitem[{{Rieke} {et~al.}(2004){Rieke}, {Young}, {Engelbracht}, {Kelly},
  {Low}, {Haller}, {Beeman}, {Gordon}, {Stansberry}, {Misselt}, {Cadien},
  {Morrison}, {Rivlis}, {Latter}, {Noriega-Crespo}, {Padgett}, {Stapelfeldt},
  {Hines}, {Egami}, {Muzerolle}, {Alonso-Herrero}, {Blaylock}, {Dole}, {Hinz},
  {Le Floc'h}, {Papovich}, {P{\'e}rez-Gonz{\'a}lez}, {Smith}, {Su}, {Bennett},
  {Frayer}, {Henderson}, {Lu}, {Masci}, {Pesenson}, {Rebull}, {Rho}, {Keene},
  {Stolovy}, {Wachter}, {Wheaton}, {Werner}, \& {Richards}}]{Rieke.etal:04}
{Rieke}, G.~H., {et~al.} 2004, \apjs, 154, 25

\bibitem[{{Rodono} {et~al.}(1986){Rodono}, {Cutispoto}, {Pazzani}, {Catalano},
  {Byrne}, {Doyle}, {Butler}, {Andrews}, {Blanco}, {Marilli}, {Linsky},
  {Scaltriti}, {Busso}, {Cellino}, {Hopkins}, {Okazaki}, {Hayashi}, {Zeilik},
  {Helston}, {Henson}, {Smith}, \& {Simon}}]{Rodono.etal:86}
{Rodono}, M., {et~al.} 1986, \aap, 165, 135

\bibitem[{{Rodono} {et~al.}(1998){Rodono}, {Pagano}, {Cutispoto}, {Marino},
  {Messina}, {Leto}, {Trigilio}, {Umana}, \& {Neri}}]{Rodono.etal:98}
{Rodono}, M., {et~al.} 1998, in Astronomical Society of the Pacific Conference
  Series, Vol. 154, Cool Stars, Stellar Systems, and the Sun, ed. R.~A.
  {Donahue} \& J.~A. {Bookbinder}, 1446--+

\bibitem[{{Scaltriti} {et~al.}(1993){Scaltriti}, {Busso}, {Ferrari-Toniolo},
  {Origlia}, {Persi}, {Robberto}, \& {Silvestro}}]{Scaltriti.etal:93}
{Scaltriti}, F., {Busso}, M., {Ferrari-Toniolo}, M., {Origlia}, L., {Persi},
  P., {Robberto}, M., \& {Silvestro}, G. 1993, \mnras, 264, 5

\bibitem[{{Schuster} {et~al.}(2006){Schuster}, {Marengo}, \&
  {Patten}}]{Schuster.etal:06}
{Schuster}, M.~T., {Marengo}, M., \& {Patten}, B.~M. 2006, in Society of
  Photo-Optical Instrumentation Engineers (SPIE) Conference Series, Vol. 6270,
  Society of Photo-Optical Instrumentation Engineers (SPIE) Conference Series

\bibitem[{{Song} {et~al.}(2005){Song}, {Zuckerman}, {Weinberger}, \&
  {Becklin}}]{Song.etal:05}
{Song}, I., {Zuckerman}, B., {Weinberger}, A.~J., \& {Becklin}, E.~E. 2005,
  \nat, 436, 363

\bibitem[{{Stark} \& {Kuchner}(2008)}]{Stark.Kuchner:08}
{Stark}, C.~C., \& {Kuchner}, M.~J. 2008, \apj, 686, 637

\bibitem[{{Strassmeier} {et~al.}(1993){Strassmeier}, {Hall}, {Fekel}, \&
  {Scheck}}]{Strassmeier.etal:93}
{Strassmeier}, K.~G., {Hall}, D.~S., {Fekel}, F.~C., \& {Scheck}, M. 1993,
  \aaps, 100, 173

\bibitem[{{Verma} {et~al.}(1983){Verma}, {Ghosh}, {Iyengar}, {Tandon},
  {Daniel}, {Sanwal}, \& {Rengarajan}}]{Verma.etal:83}
{Verma}, R.~P., {Ghosh}, S.~K., {Iyengar}, K.~V.~K., {Tandon}, S.~N., {Daniel},
  R.~R., {Sanwal}, N.~B., \& {Rengarajan}, T.~N. 1983, \apss, 97, 161

\bibitem[{{Weinberger}(2008)}]{Weinberger:08}
{Weinberger}, A.~J. 2008, \apjl, 679, L41

\bibitem[{{Werner} {et~al.}(2004){Werner}, {Uchida}, {Sellgren}, {Marengo},
  {Gordon}, {Morris}, {Houck}, \& {Stansberry}}]{Werner.etal:04}
{Werner}, M.~W., {Uchida}, K.~I., {Sellgren}, K., {Marengo}, M., {Gordon},
  K.~D., {Morris}, P.~W., {Houck}, J.~R., \& {Stansberry}, J.~A. 2004, \apjs,
  154, 309

\bibitem[{{Wilner} {et~al.}(2003){Wilner}, {Dowell}, {Holman}, \&
  {Kuchner}}]{Wilner.etal:03}
{Wilner}, D.~J., {Dowell}, C.~D., {Holman}, M.~J., \& {Kuchner}, M.~J. 2003, in
  Bulletin of the American Astronomical Society, Vol.~35, Bulletin of the
  American Astronomical Society, 1415--+

\bibitem[{{Wood} {et~al.}(2005){Wood}, {M{\"u}ller}, {Zank}, {Linsky}, \&
  {Redfield}}]{Wood.etal:05}
{Wood}, B.~E., {M{\"u}ller}, H.-R., {Zank}, G.~P., {Linsky}, J.~L., \&
  {Redfield}, S. 2005, \apjl, 628, L143

\bibitem[{{Wright} \& {Barlow}(1975)}]{Wright.Barlow:75}
{Wright}, A.~E., \& {Barlow}, M.~J. 1975, \mnras, 170, 41

\bibitem[{{Wyatt} \& {Whipple}(1950)}]{Wyatt.Whipple:50}
{Wyatt}, S.~P., \& {Whipple}, F.~L. 1950, \apj, 111, 134

\bibitem[{{Zuckerman} {et~al.}(2008){Zuckerman}, {Fekel}, {Williamson},
  {Henry}, \& {Muno}}]{Zuckerman.etal:08}
{Zuckerman}, B., {Fekel}, F.~C., {Williamson}, M.~H., {Henry}, G.~W., \&
  {Muno}, M.~P. 2008, \apj, 688, 1345

\end{thebibliography}

\clearpage

\begin{deluxetable}{lcccccccc}
\tablewidth{0pt}
\tabletypesize{\scriptsize}
\tablecaption{Summary of photometric data for the program stars\label{t:photo}}
\tablehead{
\multicolumn{1}{c}{} &
\multicolumn{6}{c}{MAGNITUDE}& 
\multicolumn{2}{c}{EXCESS} \\ 

\colhead{Target~~~~~~~~~} &
\colhead{3.6$\;\mu m$} &
\colhead{4.5$\;\mu m$} &
\colhead{5.8$\;\mu m$} &
\colhead{8.0$\;\mu m$} &
\colhead{24$\;\mu m$} &
\colhead{70$\;\mu m$} &
\colhead{L$_{dust}$/L$_{bol}$ [10$^{-2}$]\tablenotemark{\dagger}}&
\colhead{Prev. work}
}

\startdata
AR~Lac \tablenotemark{15,3}& 4.23$\pm$0.02 & 4.28$\pm$0.03 & 4.24$\pm$0.03 & 4.25$\pm$0.03 & 4.07$\pm$0.08 & $>3.06$ & $< 2.0$ & \nodata \\
AR~Psc \tablenotemark{1,2}& 4.82$\pm$0.03 & 4.87$\pm$0.03 & 4.83$\pm$0.03 & 4.81$\pm$0.03 & 4.69$\pm$0.05& $>3.26$& $1.5^{+0.4}_{-0.4}$ & \nodata \\
CF~Tuc \tablenotemark{3,4}&  5.38$\pm$0.03 & 5.42$\pm$0.03 & 5.41$\pm$0.04 & 5.37$\pm$0.04 & 5.20$\pm$0.05& 4.65$\pm$0.51& $< 2.3$ & Y \tablenotemark{a,b}\\
II~Peg \tablenotemark{2,5,6}&  4.53$\pm$0.03 & 4.56$\pm$0.03 & 4.54$\pm$0.04 & 4.51$\pm$0.04 & 4.28$\pm$0.04& 3.75$\pm$0.56& $1.9^{+0.5}_{-0.4}$ & Y \tablenotemark{b,c}\\
TY~Pyx \tablenotemark{7}&  5.20$\pm$0.03 & 5.27$\pm$0.03 & 5.26$\pm$0.03 & 5.26$\pm$0.04 & \nodata& \nodata& $<1.7$ & N \tablenotemark{d} \\
UV~Psc \tablenotemark{8}&  7.22$\pm$0.04 & 7.19$\pm$0.04 & 7.22$\pm$0.05 & 7.13$\pm$0.05 & 6.99$\pm$0.07&$> 4.50$& $< 1.9$ & \nodata \\
UX~Ari \tablenotemark{2,3,9}&  3.92$\pm$0.03 & 3.91$\pm$0.02 & 3.87$\pm$0.02 & 3.89$\pm$0.02 & 3.54$\pm$0.04& 3.09$\pm$0.30& $0.7^{+0.2}_{-0.2}$& Y \tablenotemark{e} - N \tablenotemark{a} \\
V471~Tau \tablenotemark{10,11}&  7.19$\pm$0.03 & 7.20$\pm$0.03 & 7.18$\pm$0.04 & 7.15$\pm$0.04 & 7.05$\pm$0.07& $>2.55$ & $<2.4$ & \nodata\\
WY~Cnc \tablenotemark{12,13}&  7.44$\pm$0.03 & 7.45$\pm$0.04 & 7.41$\pm$0.05 & 7.40$\pm$0.05 & 7.31$\pm$0.07& $>2.97$ &$< 2.2$ & Y \tablenotemark{b}\\
XY~Uma \tablenotemark{14}&  7.16$\pm$0.02 & 7.17$\pm$0.03 & 7.15$\pm$0.05 & 7.11$\pm$0.04 & 7.13$\pm$0.07& $>3.42$ &$< 1.9$ & \nodata \\

\enddata
\tablecomments{\scriptsize MIPS magnitudes from aperture photometry using 
the following source radii and sky background annuli.  $70\mu m$: $16\arcsec$ source,
18--39\arcsec\ background, aperture correction factor 2.04.  
$24\mu m$: 13\arcsec\ source ($6\arcsec$ for V471 Tau), 20--$32\arcsec$ background, aperture correction factor 1.17 (1.70 for V471 Tau).---$^\dagger$Measured dust luminosity values and $3\sigma$ upper limits ---\emph{BVRIJHK} photometry references:~$^{\rm{1}}$ \cite{Cutispoto.etal:01};~$^{\rm{2}}$ \cite{Messina.etal:08};~$^{\rm{3}}$ \cite{Busso.etal:88};~$^{\rm{4}}$ \cite{Cutispoto:91};~$^{\rm{5}}$ \cite{Scaltriti.etal:93} ;~$^{\rm{6}}$ \cite{Byrne.etal:98};~$^{\rm{7}}$ \cite{Busso.etal:87};~$^{\rm{8}}$ \cite{Kjurkchieva.etal:05};~$^{\rm{9}}$ \cite{Alekseev.Kozhevnikova:04};~$^{\rm{10}}$ \cite{Ibanoglu:89};~$^{\rm{11}}$ \cite{Ibanoglu.etal:05};~$^{\rm{12}}$ \cite{Milano.etal:86};~$^{\rm{13}}$ \cite{Heckert.etal:98};~$^{\rm{14}}$ \cite{Kjurkchieva:00}~$^{\rm{15}}$ \cite{Rodono.etal:86} ---~Previous work excesses from the literature: $^{\rm{a}}$ \cite{Busso.etal:88};~$^{\rm{b}}$ \cite{Scaltriti.etal:93};~$^{\rm{c}}$ \cite{Rodono.etal:98};~$^{\rm{d}}$ \cite{Busso.etal:87};~$^{\rm{e}}$ \cite{Verma.etal:83}.}

\normalsize

\end{deluxetable}

\clearpage

\begin{deluxetable}{lcc|cc|cc}
\tablewidth{0pt}
\tabletypesize{\scriptsize}
\tablecaption{Stellar parameters and best-fit model results\label{t:fit}}
\tablehead{
\multicolumn{3}{c}{}& 
\multicolumn{4}{c}{BEST-FIT MODELS} \\ 
\cline{4-7} \\

\multicolumn{3}{c}{}& 
\multicolumn{2}{c}{Photospheres} &
\multicolumn{2}{c}{Dust} \\

\colhead{Target} &
\colhead{Sp. Type\tablenotemark{a}} &
\colhead{P [day]\tablenotemark{a}} &
\colhead{T$_{eff}$ [K] \tablenotemark{b}} &
\colhead{R$_{2}$/R$_{1}$} &
\colhead{T$_{BB}$ [K]\tablenotemark{c}} &
\colhead{T$_{grain}$ [K]\tablenotemark{d}}  
}

\startdata

AR~Lac & G2IV/K0IV &  1.98 & 5750/4750 & 2.4 & \nodata  &  \nodata \\
AR~Psc & K1IV/G7V & 12.25 & 4500/5250 & 0.8 & 1900$^{+300}_{-300}$  & 1400$^{+250}_{-300}$  \\
CF~Tuc & G0V/K4V & 2.80 & 6500/4750 & 3.1 & \nodata  &  \nodata \\
II~Peg & K2IV/M2V &  6.71 & 4500 & \nodata & 1700$^{+250}_{-300}$ & 1200$^{+200}_{-200}$  \\
TY~Pyx & G5IV/G5IV &  3.20 & 5750/5500 & 0.6 & \nodata  &  \nodata \\
UV~Psc & G5V/K2V & 0.86  & 5500/5250 & 1.1 & \nodata  &  \nodata \\
UX~Ari & G5V/K0IV &  6.44 & 5250/4750 & 3.6 & 1300$^{+200}_{-200}$   & \nodata  \\
V471~Tau & K2V/DA &  0.52 & 5500 & \nodata & \nodata  &  \nodata \\
WY~Cnc & G5V/K0V &  0.83 & 5500/3500 & 0.8 & \nodata  &  \nodata \\
XY~Uma & G5V/K5V &  0.48 & 5750/4250 & 0.7 & \nodata  &  \nodata \\

\enddata

\tablecomments{\small$^{\rm{a}}$ Spectral types and periods from \cite{Strassmeier.etal:93} --~\small$^{\rm{b}}$ 
Based on Kurucz models; the error associated with the best-fit photospheric temperatures is $\sim \pm 250$K --~\small$^{\rm{c}}$ Blackbody dust temperature 
--~\small$^{\rm{d}}$ Grain emissivity law dust temperature (see text).
}

\normalsize

\end{deluxetable}

\clearpage

%
%
%
%
%
%
%
%
%
%

\end{document}